\documentclass[10pt,prl,aps,twocolumn,showpacs,floatfix]{revtex4-1}

\usepackage{graphicx}
\usepackage{amssymb}
\usepackage{subfigure}
\usepackage{color}
\usepackage{amsmath} 
\newcommand{\quarter}{\hbox{$\frac{1}{4}$}}
\newcommand{\be}{\begin{equation}}
\newcommand{\ee}{\end{equation}}
\newcommand{\la}{\langle}
\newcommand{\ra}{\rangle}

\begin{document}

\title{Thermal valence-bond-solid transition of quantum spins in two dimensions}

\author{Songbo Jin and Anders W. Sandvik}
\affiliation{Department of Physics, Boston University, 590 Commonwealth Avenue, Boston, Massachusetts 02215, USA}

\begin{abstract}
We study the $S=1/2$ Heisenberg ($J$) model on the two-dimensional square lattice in the presence of additional higher-order 
spin interactions ($Q$) which lead to a valence-bond-solid (VBS) ground state. Using quantum Monte Carlo simulations, we analyze 
the thermal VBS transition. We find continuously varying exponents, with the correlation-length exponent $\nu$ close to the 
Ising value for large $Q/J$ and diverging when $Q/J$ approaches the quantum-critical point (the critical temperature $T_c \rightarrow 0$). 
This is in accord with the theory of deconfined quantum-critical points, which predicts that the transition should approach a 
Kosterlitz-Thouless (KT) fixed point when $T_c \rightarrow 0^+$ (while the transition versus $Q/J$ for $T=0$ is in a different class). 
We find explicit evidence for KT physics by studying the emergence of $U(1)$ symmetry of the order parameter at $T=T_c$ 
when $T_c \to 0$.
\end{abstract}

\date{\today}

\pacs{75.10.Kt, 75.10.Jm, 75.40.Mg, 75.40.Cx}

\maketitle

The $S=1/2$ Heisenberg model on the two-dimensional (2D) square lattice can host a quantum phase transition between the standard 
N\'eel antiferromagnet (AFM) and a valence-bond-solid (VBS) ground state when other interactions are added \cite{Read89}. This 
transition between two different ordered quantum states has been the subject of a large body of work for more than 20 years
\cite{Sachdev08}. In the $J$-$Q$ model \cite{Sandvik07}, the pair exchange $J$ is supplemented by a product of two or more 
singlet-projectors on adjacent links of the lattice, with strength $Q$. For sufficiently large $Q/J$, the correlated singlets 
destroy the N\'eel order existing for small $Q/J$, leading to the VBS crystallization of ordered singlets. Unlike geometrically 
frustrated systems, on which searches for VBS states and the AFM--VBS transition were focused for a long 
time \cite{Chandra88,Dagotto89,Schulz96,Capriotti01}, the $J$-$Q$ model is amenable to large-scale quantum 
Monte Carlo (QMC) simulations \cite{Kaul12} and its AFM--VBS transition has been studied 
extensively \cite{Sandvik07,Melko08,Kaul08,Jiang08,Lou09,Kotov09,Sandvik10,Sandvik11,Banerjee11,Nishiyama12,Damle13}. Many results
indicate that the model realizes the unusual (``non-Landau'') deconfined quantum-critical (DQC) point proposed by Senthil 
et al.~\cite{Senthil0, Senthil1}, where the two order parameters both arise out of of emergent spin-$1/2$ degrees of freedom 
(spinons), which at criticality are described by a gauge-field theory; the non-compact CP$^1$ model. Other, less exotic 
scenarios within the standard Landau-Ginzburg-Wilson framework for phase transitions have also been put forward \cite{Jiang08,Kuklov08,Chen13}, 
however. 

The putative DQC points are manifestations of interesting quantum effects, due to Berry phases and emergent topological conservation laws
\cite{Senthil1,Kaul12}, that potentially are at play in many strongly-correlated quantum-matter systems. Being amenable to large-scale unbiased QMC simulations, further 
studies of the $J$-$Q$ class of  models offer opportunities to examine the DQC proposal in detail from various angles. Here we present results for the ordering 
transition of the VBS at finite temperature, discussing its universality, relationship to conformal field theory (CFT), and insights gained into the 
emergent U($1$) symmetry \cite{Senthil1} associated with the DQC point when approached at finite temperature.

{\it Universality of the VBS transition}---The square-lattice columnar VBS obtaining with the standard $J$-$Q$ model breaks $Z_4$ symmetry and, thus, 
it should also exist at finite temperature ($T>0$). Thermal 2D $Z_4$-breaking transitions normally do not have fixed critical exponents, but
belong to a universality class of CFTs with charge $c=1$ exhibiting continuously varying exponents (as a function of 
model parameters) \cite{cft1,cft2}. Realizations of these transitions include the standard XY model with a field $h\cos(4\theta_i$) for all spins 
$i$ (angles $\theta_i$) \cite{jose,pasquale}, the Ashkin-Teller model \cite{at1,at2}, and the Ising model with nearest- and next-nearest neighbor
interactions (the $J_1$-$J_2$ model) \cite{jin12,kalz12}. The deformed XY model has a critical line connecting Ising and Kosterlitz-Thouless 
(KT) fixed points \cite{kt1,kt2}, while the critical lines of the AT and $J_1$-$J_2$ models connect Ising and 4-state Potts points. It 
is then intersting to ask if any of these scenarios are realized in the $T>0$ paramegnet--VBS transitions of the $J$-$Q$ model. 
In this Letter we present strong evidence for universality corresponding to the Ising--KT critical line, with the KT transition 
obtaining in the limit when $Q/J$ approaches its quantum-critical value and the critical temperature $T_c \to 0$. This is in agreement 
with the DQC theory and its U$(1)$ gauge-field description, where the nature of the VBS state is dictated by a dangerously irrelevant 
operator \cite{Sachdev08,Senthil0,Senthil1}, which implies that the VBS fluctuations should cross over from $Z_4$ to U($1$) symmetric as the 
DQC point is approached, which in fact has been observed in ground state studies of the VBS fluctuations of $J$-$Q$ models 
\cite{Sandvik07,Jiang08,Lou09}. We here show explicitly that this also applies to the $T>0$ critcal line when $T_c \to 0$.

The $T>0$ VBS transition was previously studied by Tsukamoto, Harada and Kawashima \cite{Kawashima09}, who carried out QMC simulations of 
the $J$-$Q_2$ version of the $J$-$Q$ model, where the $Q_2$ interaction is one of products of two singlet projectors. The results were 
puzzling, with significant deviations from the ``weak universality'' scenario applying to the transitions discussed above, where the critical 
correlation-function exponent $\eta=1/4$ is constant (while other exponents depend on system details). Instead, $\eta \approx 0.5$ was
obtained \cite{Kawashima09}. Here we consider the $J$-$Q_3$ model \cite{Lou09}, where the $Q_3$ term consists of three bond-singlet projectors 
(forming columns on three adjacent lattuce links). This model has a much more robust $T=0$ VBS for large $Q_3$, while the VBS state of the $J$-$Q_2$ 
model is near-critical even for $Q_2/J \to \infty$. With the $J$-$Q_3$ model we can systematically study the $T>0$ transition both far away from 
the DQC point and close to it. We find consistency with $\eta=1/4$ to high precision, and also point out that cross-over behavior related to 
the DQC  criticality exactly at $T=0$ makes it difficult to reliably extract the exponents when $T_c$ is low. We believe that this behavior 
affected the previous study of $\eta$.

{\it Model and methods}---We next discuss the QMC calculations and data analysis on which we base our conclusions.
The $J$-$Q_3$ Hamiltonian is defined as
\be
H=-J\sum_{\la i,j\ra}P_{ij}-Q_3\sum_{\la ijklmn \ra}P_{ij}P_{kl}P_{mn},
\label{equ:hamil}
\ee
where $P_{ij}$ is a nearest-neighbor bond-singlet projector;
\be
P_{ij}=\quarter-\mathbf{S}_i\cdot\mathbf{S}_j,
\label{equ:projector}
\ee
here on the square lattice with $L^2$ sites.
We define the coupling ratio $q=Q_3/J$. The quantum-critical point separating the AFM and VBS states is $q_c = 1.500(2)$ \cite{Lou09}. 
We here use the stochastic series expansion (SSE) QMC method with loop updates \cite{sse0,sse1,sse2} to compute several quantities 
useful for extracting the critical temperature and exponents of the VBS transition for $q > q_c$.

There are various ways to define the VBS correlation length. For computational convenience we here use a definition based on the 
$J$-term (bond) susceptibility,
\be
{\chi}_{b_1,b_2}=\int_{0}^{\beta}d\tau\langle{P}_{b_2}(\tau){P}_{b_1}(0)\rangle,
\label{equ:susc1}
\ee
where $P_b$ is a singlet projector as in (\ref{equ:projector}), with $b$ denoting a bond connecting sites $i_b,j_b$.
These susceptibilities can be computed easily with the SSE method, because the bond operators are terms of the Hamiltonian and, thus, appear 
in the sampled operator sequences. With $n(b)$ denoting the number of $J$-operators on bond $b$ in the sequence, the susceptibility is 
given by \cite{Sandvik97}
\be
{\chi}_{b_1,b_2}=\langle n(b_1)n(b_2)-\delta_{b_1,b_2}n(b_1)\rangle /\beta.
\label{equ:susc2}
\ee
This estimator works well as long as $q$ is not too large. When $q>10$ the measurements become noisy due to the low density of bond operators, but
for our purposes here this is not a problem.

\begin{figure}
\center{\includegraphics[width=8cm, clip]{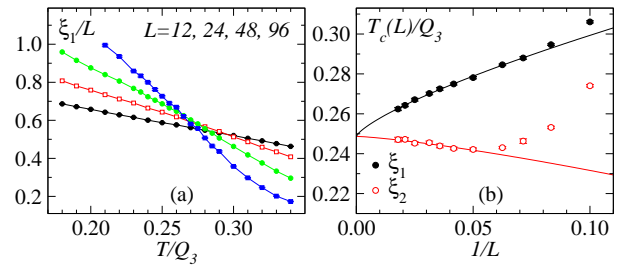}}
\vskip-2mm
\caption{(Color online) Extraction of $T_c$ for system at $q=5$. Shown in (a) are, in order of higher to lower curves on the left side,
results for $\xi_1/L$ versus $T$ for system sizes $L=96,48,24$, and $12$. Crossing points giving $T_c(L)$ estimates are shown in (b), using
both $\xi_1$ and $\xi_2$ with sizee pairs $(L,2L)$. The data were fit to the form $T_c(L)=T_c(\infty)+a/L^w$ in the range $1/L \in [0,0.08]$ ($\xi_1$) 
and $[0,0.06]$ ($\xi_2$), yelding $T_c=0.249(3)$ in the case of $\chi_1$. For the $\xi_2$ fit, $T_c(\infty)=0.249$ was fixed.} 
\label{fig1}
\vskip-2mm
\end{figure}

To detect columnar VBS order, we consider the bonds $b_1$ and $b_2$ oriented in the same ($x$ or $y$) lattice direction and denote 
by $\chi^\alpha(\mathbf{r})$, $\alpha=x,y$, the spatially averaged distance-dependent susceptibility. The VBS susceptibility $\chi^x_{\rm VBS}$ 
is the $\mathbf{q}=(\pi,0)$ Fourier transform of $\chi^x(\mathbf{r})$ (and analogously for $y$). Because the columnar VBS breaks the lattice
rotational symmetry, we can define two correlation lengths. Using the $x$ susceptibility and defining $\mathbf{q}_0=(\pi,0)$, 
$\mathbf{q}_1=(\pi+{2\pi}/{L},0)$ and $\mathbf{q}_2=(\pi,{2\pi}/{L})$ we have the correlation lengths parallel and 
perpendicular to the x-oriented bonds for an $L\times L$ lattice;
\be
\xi^x_1=\frac{L}{2\pi}\sqrt{\frac{\chi^x_{\rm VBS}(\mathbf{q}_0)}{\chi^x_{\rm VBS}(\mathbf{q}_1)}-1},~~~
\xi^x_2=\frac{L}{2\pi}\sqrt{\frac{\chi^x_{\rm VBS}(\mathbf{q}_0)}{\chi^x_{\rm VBS}(\mathbf{q}_2)}-1},
\label{xi}
\ee
and analogously for $y$. Average valuess of $x$, $y$ quantities are denoted in the
following without superscript.

{\it Critical temperature}---To illustrate how the critical VBS temperature $T_c$ is determined, Fig.~\ref{fig1}(a) shows $\xi_1/L$ versus $T$ at $q=5$ 
for several system sizes. According to standard finite-size scaling theory \cite{Fisher}, $\xi_1/L$ for different $L$ should cross at $T_c$
when $L \to \infty$. Due to expected scaling corrections, the crossing point $T_c(L_1,L_2)$ between two system sizes, which we here take as $L$ and $2L$, 
drifts slowly with $L$ and converges as the system size increases. We use the crossing point for both $\xi_1$ and $\xi_2$ to extract $T_c$ and 
check the consistency of the two results.

Fig.~\ref{fig1}(b) shows two sets of $T_c(L)$ point obtaied from $\xi_1$ and $\xi_2$. Both curves can be fitted with 
the form $T_c(L)=T_c(\infty)+a/L^w$ but the parameters are different. The two curves appoach $T_c$ from different directions. 
The $\xi_1$ data have large deviations from the fitted function only for small systems ($L \alt 12$), while $\xi_2$ shows corrections 
extending up to larger systems and the size dependence is non-monotonic. In spite of the different behaviors, the data 
extrapolate consistently to a common $T_c$ in the thermodynamic limit. To demonstrate this, we show in Fig.~\ref{fig1}(b) a 
fit to the $\xi_1$ data, which gives $T_c=0.249(3)$. (which has a smaller statistical error than the value from $\xi_2$). We also show a 
fit to the  $\xi_2$ data, where the $T_c(\infty)$ value is fixed at the result based on $\xi_1$. 

\begin{figure}
\center{\includegraphics[width=8cm, clip]{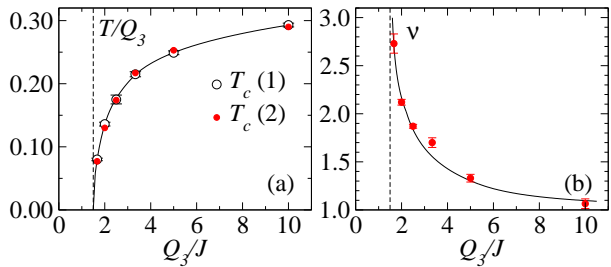}}
\vskip-2mm
\caption{(Color online) (a) The critical temperature extracted from $\xi_1/T$ (open circles). 
Also shown are results (solid circles) where the VBS susceptibility exhibits the best scaling behavior when $\gamma=7/4$ 
is fixed. (b) The exponent $\nu$ versus $q$. The vertical dashed lines in both panels mark 
the quantum-critical ratio $q_c$ \cite{Lou09}. The curves are guides to the eye.}
\label{fig2}
\vskip-4mm
\end{figure}

$T_c$ values for several other $q$ points were extracted in the same way, making sure that $\xi_1$ and $\xi_2$ data extrapolate consistently
but using only the $\xi_1$ results (which always have smaller errors) for further analysis. This procedure becomes increasingly challenging as the 
quantum-critical point $q_c$ is approached and $T_c \to 0$. The corrections to the asymptotic form became more profound and larger systems 
have to be used. In addition, the SSE calculations become more time-consuming, since $L \gg  1/T$ is required for the simulated
effective classical system to be firmly in the 2D limit. The largest system simulated was $L=192$ at $q={5}/{3}$. Results for $T_c$ 
are shown versus the coupling ratio in Fig.~\ref{fig2}(a).

\begin{figure}
\center{\includegraphics[width=8cm, clip]{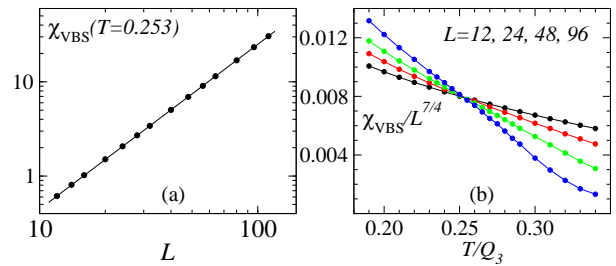}}
\vskip-2mm
\caption{(Color online) (a) Scaling behavior of the critical VBS susceptibility for systems at $q=5$. Here $T$ was
adjusted to give the best linear scaling on the log-log plot, giving $\gamma/\nu=1.750(1)$. (b) The size-scaled
susceptibility under the assumption $\eta=1/4$ versus $T$ for several system sizes. The crossing point is consistent with 
$T_c$ extracted from the correlation length.}
\label{fig3}
\vskip-2mm
\end{figure}

{\it Critical exponents}---we next present an analysis of the scaling behavior of the VBS susceptibility, which exactly at $T_c$
should follow the form
\be
\chi^{}_{\rm VBS}(T_c) \sim L^{\gamma/\nu},
\label{scaling}
\ee
where $\gamma/\nu = 2 - \eta$. Here we can use the value of $T_c$ extracted above from the correlation length scaling. Alternatively, we can
adjust the temperature until the best power-law scaling is obtained. If sufficiently large system sizes are used the two methods should of
course deliver consistent results. This is indeed the case, as shown in Fig.~\ref{fig2}(a). An example of the best power-law scaling is shown 
for the system with $q=5$ in Fig.~\ref{fig3}(a). Here the corrections to scaling appear to be very small (i.e., a straight line can be well fitted on 
the log-log scale even when systems as small as $L=10$ are included) and the temperature, $T=0.253$, is only about one error bar off the $T_c$
value extracted from $\xi_1/L$. A series of fits with a bootstrap analysis to estimate the errors yielded $\gamma/\nu=1.750(1)$, corresponding 
$\eta=0.250(1)$. Thus, we find complete consistency, to rather high precision, with the most natural expectation of $\eta=1/4$. We obtain 
similar results for all values of $Q_3/J$ studied.

Fig.~\ref{fig3}(b) demonstrates a different way to analyze the susceptibility and test the assumption $\eta=1/4$, by graphing $\chi_{\rm VBS}L^{-{7}/{4}}$ 
versus $T$ is for different system sizes. All curves cross essentially at the same point, which confirms the scaling power $\gamma/\nu={7}/{4}$ 
in Eq.~(\ref{scaling}). The remarkable absence of drift in the crossing points of $\chi_{\rm VBS}L^{-{7}/{4}}$ (in contrast to the significant
drift found for the normalized correlations lengths) makes this quantity a perfect candidate for carrying out a finite-size data collapse to extract 
correlation length exponent $\nu$, which we consider next.

\begin{figure}
\center{\includegraphics[width=6.5cm, clip]{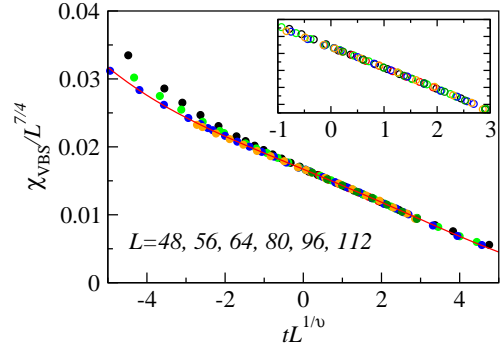}}
\vskip-2mm
\caption{(Color online) Data collapse of the VBS susceptibility for system s at $ q={10}/{3}$. The inset shows data for $L=80, 96, 112$ 
in the range $tL^{1/\nu} \in [-0.5,3]$ for which the fitting procedure was carried out. The main part shows data in a larger window and 
including also smaller systems. The fit yelded $\nu=1.70(5)$.}
\label{fig4}
\vskip-4mm
\end{figure}

Shown in Fig.~\ref{fig4} are data sets for system sizes $L=48$ to $112$ at $q={10}/{3}$, graphed versus $tL^{1/\nu}$, where $t$
is the reduced temperature, $t = (T-T_c)/T_c$, and the critical temperature was determined in the manner above to be $T_c=0.217$. The
correlation lengt $\nu$ was adjusted to give the best data collapse, as measured with respect to a  polynomial fitted simultaneously to all 
data points for $L=80,96,112$ in the range  $tL^{1/\nu} \in [-0.5,3]$. A zoom-in on this window is shown in the inset. The fit was restricted
to the larger sizes in order to minimize the effects of neglected scaling corrections, and the window of $tL^{1/\nu}$ values was chosen to 
obtain a statistically sound fit. This procedure along with an analysis of the statistical errors gave $\nu=1.70(5)$. When $q$ 
is tuned towards $q_c$, larger system sizes are required to achieve good collapse due to more pronounced scaling corrections, as already
mentioned above. As an example, at $q={5}/{3}$, we used system sizes $L=112, 128, 160, 192$. 

All our results for $T_c$ and $\nu$ versus $q$ are shown in Fig.~\ref{fig2}. $T_c$ clearly decreases when $q$ approaches $q_c$ and $\nu$ grows rapidly, 
changing from $1.065(5)$ at $q=10$ to $2.7(1)$ at $q={5}/{3}$. The behavior suggests that $\nu$ diverges when $q\to q_c$, which would mean that
the critical line corresponds to the $c=1$ Ising--KT scenario, with the KT universality applying in the limit $q\to q_c^+$ and 2D Ising
universality ($\nu=1$) applying in the extreme limit far from the quantum-critical point (which cannot strictly be achieved within the $J$-$Q_3$ model,
but $\nu$ is already close to the Ising value for $q=10$; the largest $q$ studied here). This scenario is also supported by the fact that there
is no specific-heat peak at $T_c$, i.e., the exponent $\alpha <0$.

\begin{figure}
\center{\includegraphics[width=5.5cm, clip]{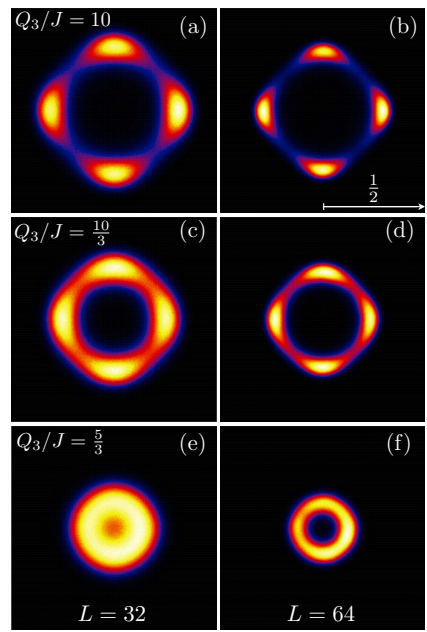}}
\vskip-2mm
\caption{(Color online) Dimer-order distribution $P(D_x,D_y)$ for system size $L=32$ (left panels) and $L=64$ (right panels)
in the close vicinity of $T_c$. The coupling ratios (temperatures) are $q=10$ ($T=0.29$) in (a),(b);
$q={10}/{3}$ ($T=0.218$) (c),(d); $q={5}/{3}$ ($T=0.08$) in (e),(f). In (f) the distributions
is effected by unequal sampling (due to long QMC autocorrelation times) in different angular sectors.}
\label{fig5}
\vskip-4mm
\end{figure}

{\it Emergent U(1) symmetry}---The changing critical exponents are related to an evolution of the critical VBS fluctuations. We investigate these
by following the distribution of the components $(D_x,D_y)$ of the VBS order parameter. The columnar VBS operator for x-direction bonds are defined as
\be
\hat D_x=\frac{1}{N}\sum_{\mathbf{r}}(-1)^{x}P_{\mathbf{r},\mathbf{r+\hat x}},
\label{equ:dx}
\ee
and $\hat D_y$ is defined analogously. An SSE-sampled  configuration can be assigned definite ``measured'' values $(D_x,D_y)$
by the operator-counting procedure discussed above in the context of the susceptibility (\ref{equ:susc1}). We accumulate the probability distribution
$P(D_x,D_y)$, which reflects the nature of the VBS fluctuations. In analogy with XY models with
dangerously-irrelevant $Z_4$ perturbations \cite{Lou07}, one would expect the four-fold symmetric VBS distribution to develop signatures
of U($1$) symmetry. This has previously been observed when approaching the quantum-critical point at $T=0$. We now approach this
point by following the $T>0$ critical line. Fig.~\ref{fig5} shows results for several combinations of the system size and the coupling ratio.
While clearly four-fold symmetric distributions apply for large $q$, the histograms become more circular-symmetric as the quantum-critical 
point is approached. As at $T=0$ \cite{Lou09}, one would expect the distribution to be effectively U($1$) symmetric when $L$ (or some other the course-graining 
scale) is less than a lengt-scale $\Lambda$, with $\Lambda \to \infty$ as $q\to q_c$. For the system sizes studied, we are below $\Lambda$ at $q=5/3$, while
for the larger $q$ in Fig.~\ref{fig5} the system sizes exceed $\Lambda$. These observations provides direct evidence for U($1$)-symmetric VBS fluctuations 
leading to the large $\nu$ found here close to $q_c$.

{\it Discussion}---All our calculations show consistently that the thermal VBS transition in the $J$-$Q_3$ model has critical exponents 
varying in a range expected in a particular subclass of $c=1$ CFTs. The exponent $\eta$ is constant at $\eta=1/4$, in agreement with weak 
universality, and $\nu$ grows rapidly as the quantum-critical point is approached, indicating an emergent U($1$) symmetry of the VBS order 
parameter and a KT transition obtaining in the limit $T_c \to 0^+$. While we cannot strictly rule out a change of behavior to a first-order transition for very 
low temperatures \cite{Jiang08,Sandvik10,Chen13}, there are no indications of this in any of our results. Note, in particular, that in finite-size 
scaling at a first-order transition one should see $\nu=1/d$ \cite{Vollmayr}, where $d$ is the dimensionality (i.e., $d=2$ in our case when $T_c>0$). 
Instead, at the lowest $T_c$ reached here, $\nu \approx 3$. We expect that the same behavior should apply also in the $J$-$Q_2$ model, but that cross-over 
behaviors associated with the proximity to the quantum-critical point for all $Q_2/J$ in that model may make it difficult to extract the exponents there
\cite{Kawashima09}.

The significance of establishing the nature of the $T>0$ critical line is that it puts the phase diagram of the
$J$-$Q$ model firmly within an established CFT. In the limit $T \to 0^+$, the effective $(2+1)$-dimensional system, obtained
in a quantum--classical mapping through the path integral, can still be considered finite in the ``time'' dimensions, and,
thus, the KT scenario can apply. Exactly at $T=0$ the effective system is fully 3D and a different criticality must apply (that of the 
putative DQC point). Since microscopic details should not matter, by universality our results should apply to a wide range of VBSs. 

The non-commutability of the limits $L \to \infty$ and $1/T \to \infty$ is also associated with interesting
cross-overs, which we have observed here but not studied in detail. Further investigations of this aspect of the AFM--VBS 
transition are warranted.

{\it Acknowledgments---}This research was supported by the NSF under Grant No.~DMR-1104708.

\null

\end{document}